\def\revtex{1}

\ifx\revtex\undefined

\documentclass[entropy,article,submit,oneauthor,pdftex]{Definitions/mdpi}

\firstpage{1}
\pubvolume{1}
\issuenum{1}
\articlenumber{0}
\pubyear{2021}
\copyrightyear{2021}
\datereceived{}
\dateaccepted{}
\datepublished{}
\hreflink{https://doi.org/} 

\Title{The Mosquito incident and the need-(not-)to-know}

\TitleCitation{The Mosquito incident and the need-(not-)to-know}

\Author{Karl Svozil $^{1}$\orcidA{}}

\AuthorNames{Karl Svozil}

\AuthorCitation{Svozil, K.}

\address[1]{%
$^{1}$ \quad Institute for Theoretical Physics, TU Wien, Wiedner Hauptstrasse 8-10/136, 1040 Vienna,  Austria; svozil@tuwien.ac.at; \url{http://tph.tuwien.ac.at/~svozil}}

\corres{Correspondence: svozil@tuwien.ac.at}

\abstract{I am reporting here how professor Cristian Sorin Calude, henceforth called Cris, became involved in the ``Mosquito incident''. More generally, the relationship between the single individual vis-\`a-vis the state or collective is reviewed, with special emphasis to need-(not)-to-know secrets.}

\DeclareFontFamily{U}{bbold}{}
\DeclareFontShape{U}{bbold}{m}{n}
 {
  <-5.5> s*[1.069] bbold5
  <5.5-6.5> s*[1.069] bbold6
  <6.5-7.5> s*[1.069] bbold7
  <7.5-8.5> s*[1.069] bbold8
  <8.5-9.5> s*[1.069] bbold9
  <9.5-11> s*[1.069] bbold10
  <11-15> s*[1.069] bbold12
  <15-> s*[1.069] bbold17
 }{}

\begin{document}

\else
\documentclass[%
      reprint,
   twocolumn,
 amsmath,amssymb,
 aps,
 pra,
  longbibliography,
 ]{revtex4-2}

\usepackage[dvipsnames]{xcolor}

\usepackage{mathptmx}

\usepackage{amssymb,amsthm,amsmath,bm}

\usepackage{tikz}
\usetikzlibrary{calc,decorations.pathreplacing,decorations.markings,positioning,shapes,snakes}

\usepackage[breaklinks=true,colorlinks=true,anchorcolor=blue,citecolor=blue,filecolor=blue,menucolor=blue,pagecolor=blue,urlcolor=blue,linkcolor=blue]{hyperref}
\usepackage{graphicx}
\usepackage{url}

\ifxetex
%
%
\usepackage{fontspec}
\usepackage{fontspec}
\setmainfont{Garamond}
\setsansfont{Garamond}
\fi

\usepackage{mathbbol} 

\begin{document}

\title{The Mosquito incident and the need-(not-)to-know}
\thanks{This historic deliberation is a contribution to a collection entitled {\it ``Liber Amicorum
Cristian S. Calude 70''}, edited by Cezar Campeanu, Michael J. Dinneen and the author,
on the occasion of Professor Calude's 70th birthday, pages 73-80 (2022); available from \url{http://grail.smcs.upei.ca/LACris70/LiberAmicorumCris70.pdf}}

\author{Karl Svozil}
\email{svozil@tuwien.ac.at}
\homepage{http://tph.tuwien.ac.at/~svozil}

\affiliation{Institute for Theoretical Physics,
TU Wien,
Wiedner Hauptstrasse 8-10/136,
1040 Vienna,  Austria}

\date{\today}

\begin{abstract}
I am reporting here how professor Cristian Sorin Calude, henceforth called Cris, became involved in the ``Mosquito incident''. More generally, the relationship between the single individual vis-\`a-vis the state or collective is reviewed, with special emphasis to need-(not)-to-know secrets.
\end{abstract}


\maketitle

At some point in history, Cris lived in ideological tyranny called ``communism''.
Economically, communism brought austerity, frustration and devastation.
Because due to its collectivist economic principles that suppressed individualistic entrepreneurship
this totalitarian form of government
was unable to provide consumerism and the type of cargo cult which,
due to human technologic and energetic advances resulting in huge productivity increases,
flourished elsewhere.
Communism wasted huge amounts of resources to produce very little goodies consumable by its unfortunate citizenry.

It is my sombre observation that history provides ample evidence that more common equality results in more common misery.
Conversely, steepening inequalities---the rich get richer---result in improved living conditions of the poor.
Such is the result of compound interest; and, besides catastrophes---some of them economical like communism---and war,
there are hardly any means to bring down inequality~\cite{Scheidel-2017}.
The late Dirac's observation on the futility of war~\cite{dirac-81} adds an impasse by pointing out that there is,
in modern terminology, a dilemma between equality of opportunity and motivation for outcome:
for, Dirac observed, there are two equally justified socio-political positions---the right of parents to pass on their achievements to their children,
as well as the right of every child for equal chances of well-being---that shan't be the topic negotiated by war among conflicting ideologies.

Because of the misery it had instigated, communism had to curb and control the flow of information and people.
For it won't be for long that, in a more liberal unrestricted environment, the most gifted and able
persons escape to other places where life was better---a kind of osmosis.
Who can be blamed for that?

These, I kindly remind you, were the times of the ``iron curtain'' and the ``Berlin wall'',
times of heroes and villains---recall
John F. Kennedy's boyish-pathetic ``ick bin ein Berlinner!'', and his manly stance in breaking the
Soviet blockade of West Berlin with a massive airlift.
From the communistic inside it was rather obvious who was who; in particular who was the villain---no doubt about that.
People suffered through this communist tyranny on a day-to-day basis; their misery was just too obvious and evident.
But, of course, it was also the time of Dwight D. Eisenhower's epic ``good luck with our military industrial complex'' farewell speech.

Coming back to the story, in those times Cris corresponded with a colleague from one of the United States of America's ``outremer'' colonies,
aka ``the West''.
And, as noted by Cristian Terhes, a Romanian member of the European Parliament in an October 28, 2021
press conference about a heavily redacted contract
between the European Commission and a big pharmaceutical company~\cite{EC-Pfizer-contract}:
``the difference between tyranny and democracy is very simple:
when the government knows everything about you, that's tyranny.
I know how it is to live in tyranny.
When you know everything about your government, that's democracy.''
Accordingly, the Romanian government at the time wanted to know everything about the correspondence between
Cris and the West; in particular not only the metadata, but also the content.
In those days the easiest method for this to achieve was to physically open the correspondence.
Knowing this, Cris, therefore, devised an ingenious way of reflection on this eavesdropping:
he would write in his letter to a western colleague that it can be easily established if someone had opened and read the letter.
For if, upon its delivery, no mosquito could be found inside of the envelope, this
would clearly indicate that the letter had been opened, its content compromised, and then
sealed and resubmitted before its final delivery.

An easy and at the same time effective way of certification---at first sight. And sure enough, when the colleague in ``the West'' opened the letter~$\ldots $ out fell a mosquito.

Only Cris had never deposited a mosquito in the letter in the first place!

In that way, the legit sending and receiving parties had, beyond doubt, ascertained that the Securitate---the popular term for the Romanian Departamentul Securitatii Statului
(Department of State Security)---had opened, read, and understood the latter in its immediate meaning---and had put in a mosquito upon closing it.
It is still is a sarcastic pleasure to imagine those malign bureaucrats at
Securitate---maybe feaverishly and desperately---catching a mosquito, and inserting it into the letter,
in their preparation for re-sending it to ``the West''.
Little did they know that Cris had arranged with his colleague this to be a token of authentication unbeknownst to them!
As a mathematician you could also say that they made Securitate perform an involutary Cantor diagonalization on parts of the message---metacryptography
of sorts.

As far as I know, this was not a single ``mosquito incidence'' but was applied to each letter of a series of letters,
each one containing a properly apprehended mosquito upon delivery.
Poor little buggers!

Cris told me that this type of tampering with the powers that be was not entirely uncommon.
Indeeed, mathematicians tried various methods to trick the regime without being punished.
In this case, Securitate didn't suspect any malicious play and nobody stopped the letters---probably
because they thought the outcome might have been worse for them.
After all, they lived in that same hermetic tyranny, which threatened all of them.

Other instances involved an internationally well-known political opponent of the regime who was the youngest doctor in mathematics
in the country for some time before Cris got his Ph.D. a few months younger in 1976.
Cris told me that this person played many intelligent tricks.
For example, he knew and told lots of political jokes.
One could go to jail for spreading, even for listening to some such jokes.
He enjoyed telling them to the big political cats in the university.
So these guys tried to avoid him, sometimes in hilarious attempts, because these
were no-no situations: on the one hand, listening to them might have turned out to be dangerous;
and yet, on the other hand, not listening was kind of embarrassing: well, you cannot even
enjoy a joke, mate!

Cris told me that from this person he learned the following dictum: ``only those who work make mistakes''.
(My maybe incorrect reading of this is: those in power are those who work very hard and thereby make huge mistakes.)
To understand
its context, one needs to know that men in communism
could be convicted to jail if they did not work. Such were these times!

As already pointed out earlier, my own judgements about these tyrannies coincided with those who had to bear and suffer
through them; the misery and suppression was just to obvious and totally unsophisticated.
And yet, I may have been able to afford more nuanced views about western governments.
For instance, during my visits to the late Soviet Union in 1986 and 1987, it was obvious that
state tyranny was prevalent---evil was easily identifiable, and was commonly acknowledged.
For instance, the first thing the Soviet colleague taking care of me did was to direct me to a little wooded area on the banks of the
Moskva river, very close to the Lomonosov Moscow State University, and asked me if I was a member of a communist party.
Upon my negation and curiosity, he told me that he just had to make sure that I am no communist cadre sent to spy on him.
Caution demanded that he had to be very careful and kept his distance otherwise.
I guess the physicists there would have encapsulated me like a tumor, refusing to cooperate.
I should point out that the Soviet Russian colleagues I met and cooperated turned out to be the most kind, friendly
and cooperative persons I had met thus far. They considered a copy machine as the ``most sexiest thing'' in town.
Maybe I did not get the joke.

Alas, those ``in the West'' may find it much more difficult to cope with their governments when it comes to the treatment of its citizens.
There was no outright, easily acknowledged evil; and ``the West'' fed its citizens well by implementing a cargo cult-type consumerism.
We indulged ourselves in conveniences, travel, and relative freedom of speech---as long as we were not influential
and did not stir too much attention.
I had even co-organized peace rallies in Vienna before I joined the Department of Energy's Lawrence Berkeley Laboratory as a visiting scholar.
But I wonder how much autonomy we really had in all of this.
This relates to Zizek's article on WikiPedia mentioned later~\cite{ZiZek-Wiki}.

There seems to be a prevailing dichotomy between governments and the collective on the one side vis-\`a-vis its individual constituents on the other side.
It would be much to easy---and this has been brought forward by many historic thinkers such as Confuzius, Plato, and Rousseau---to assume that
governments are ``by the people and for the people'', and that its subjects should therefore obey its hopefully benign and merciful and wise rulers.
Indeed, there exists a long tradition of perceiving rulers and government as potentially malign if not outright bad and dangerous.

Again and again, at issue is the concept of the individual human being vis-\`a-vis the collective, the state and its apparatus.
We find an individual stance, a will, already in the epic of Gilgamesh~\cite{Gilgamesh-AndrewGeorge}.
Athenians in their short classical democratic period tried to curb government by introducing sortition---that is, elements of the random selection of offices,
as manifested in the kleroterion, an early randomness generator~\cite{dow_aristotlekleroteria_1939}---as well as ostracism.
The British ``Magna Carta Libertatum'' commonly called Magna Carta, the United States Constitution with all its Amendments,
and the Federalist Papers, as well as Montesquieu's Separation of Powers
were inspired by individual rights and freedom vis-\`a-vis the collective.
Socrates may have thought of this too.
In more recent times Popper~\cite{Popper-TOSAIE} as well as  Berlin~\cite{berlin-liberty}, Shafarevich~\cite{Shafarevich-tsp}
and Ayn Rand preferred the primate of individualism with regard to its freedom of expression, self-determination, and autonomy.

And yet, how far should autonomy go? Is there a mandatory, citizen's inalienable rights to know;
a desire to know the facts to evaluate a situation that cannot be transferred to others, even by legal procedures and ``oversight'' by
state functionaries and representatives?
How extended needs the freedom of information, the people's right to know, be? Is there any reason
for compartmentalization and restriction by ``need-to-know''?
And are there ``outer'' restrictions to ``need-to-know'' that curb autonomy?

It is generally assumed that for obvious reasons certain truths---e.g., so-called ``state secrets''---cannot be revealed to adversaries.
Ernst Specker once quoted similarities with ``Jesuit lies''; the sort of allowed ``white lies'' to (allegedly) prevent greater harm than the lie itself.
If, for instance, you told an adversary your line of defense or attack, this would greatly increase his ability to overcome your plans
and defeat you. Some of this is reflected in Popper's ``paradox of tolerance''.
Zizek partially agrees and puts forward some anecdotal evidence for nondisclosure also when it comes to politeness and tact
that could seriously affect serious politics~\cite{ZiZek-Wiki};
yet he also mentions that people's right to know includes revelations of the WikiLeaks story.
He speaks of ``moments of crisis for the hegemonic discourse---when one should take the risk of provoking the disintegration of
appearances'', with all the repercussions this might entail.
Ellsberg's publication of the Pentagon Papers and other secret material published by Snowden might fall in this category.

So what exactly needs to be kept secret and what has to be disclosed? This is not a question of black-and-white but nuance.
Imagine, for instance, it would be revealed
that our planet is visited by various alien species and life forms, in particular
ones resembling humans, as well as insectoids and reptilians, some of them telepathic;
that these life-forms can move in crafts that are unimpeded by our military;
that the only time we unintentionally shot down a craft of the ``others'' was on July 9, 1962, as operation
Starfish Prime produced a massive electromagnetic pulse
from a nuclear explosion at an altitude of 250 miles above the Pacific Ocean;
that the ``others'' intentions are not always benign, as they abduct humans, mutilate cattle and cause nuclear weapon systems to engage and
stop functioning; that even the head of the United Nations, Javier P\'erez de Cu\'ellar, was ``ordered'' by them
not to disclose their existence~$\ldots$ suppose all of this is
true~\cite{DeLongeHartley-SM1,DeLongeLevenda-Gods,DeLongeHartley-SM2,DeLongeLevenda-Men,Dolan-alien-agendas,Lacatski-2021}---then
what---``how about that''~\cite{UFO56}?

Would it not be better to hide these inconvenient truths from the public,
and maintain a facade like in the 1998  movie ``The Truman Show''~\cite{TruemanShow}?
I would like to raise this question, as expressed in the
Japanese-Polish movie ``Avalon''~\cite{Avalon2001}:
``What is the better game, one of which you think you can leave but can't,
or one that looks impossible to leave but an exit always exists?''
How much need-to-know if that knowledge is harmful or disillusioning?
And yet, as noted by Goethe, ``No one is more a slave than he who thinks he is free without being so.''

Maybe truth is a relative concept depending on context and means, an multi-layered and sometimes even amorphous entity.
It varies with respect to the means and purposes available; almost appearing as an epistemic concept.
I am well aware that such a stance disputes the ``ontic'' and ``pathetic''
conception of truth by, for instance,
Hannah Arendt, as ``conceptually, we may call truth what we cannot change''~\cite{Arendt-1967-truth}.
For we have no absolute means, no metaphysical Archimedean anchor, to establish this kind of ``truth that we cannot change''.
Let alone the situation that, in its extreme form, we have no autonomy to commit ourselves to such truth or facts.
And we may not even be allowed, neither might we desire, to acknowledge that on a meta-level,
so that we may not even want to know what we are not allowed to know.
And yet it sometimes helps to keep in mind the unknown unknowns.

\fi

\ifx\revtex\undefined

\funding{
}


\conflictsofinterest{The author declares no conflict of interest.
The funders had no role in the design of the study; in the collection, analyses, or interpretation of data; in the writing of the manuscript, or in the decision to publish the~results.}

\else





\fi

\ifx\revtex\undefined

\end{paracol}
\reftitle{References}


 \externalbibliography{yes}
 \bibliography{svozil}

\else


%

\fi
\end{document}